\documentclass[aps,prl,twocolumn,superscriptaddress]{revtex4-1}

\usepackage{amsmath,amssymb,color,graphicx,mathpazo,mathtools,times}
\usepackage[colorlinks=true,allcolors=blue]{hyperref}
\usepackage{soul}

\newcommand{\erefs}[1]{Eqs.~\eqref{eq:#1}}
\newcommand{\Fref}[1]{Figure~\ref{fig:#1}}
\newcommand{\fref}[1]{Fig.~\ref{fig:#1}}

\begin{document}

\title{Chimera states in small disordered optomechanical arrays}

\author{Karl Pelka}
\email{karl.a.pelka@um.edu.mt}
\affiliation{Department of Physics, University of Malta, Msida MSD 2080, Malta}
\author{Vittorio Peano}
\affiliation{Max Planck Institute for the Science of Light, Staudtstra{\ss}e 2, 91058 Erlangen, Germany}
\affiliation{Department of Physics, University of Malta, Msida MSD 2080, Malta}
\author{Andr\'e Xuereb}
\affiliation{Department of Physics, University of Malta, Msida MSD 2080, Malta}

\date{\today}

\begin{abstract}
Synchronization of weakly-coupled non-linear oscillators is a ubiquitous phenomenon that has been observed across the natural sciences. We study the dynamics of optomechanical arrays---networks of mechanically compliant structures that interact with the radiation pressure force---which are driven to self-oscillation. These systems offer a convenient platform to study synchronization phenomena and have potential technological applications. We demonstrate that this system supports the existence of long-lived chimera states, where parts of the array synchronize whilst others do not. Through a combined numerical and analytical analysis we show that these chimera states can only emerge in the presence of disorder.
\end{abstract}

\maketitle
\emph{Introduction.}---The synchronization of weakly-coupled oscillators is a common feature of non-linear dynamics that arises in various disciplines ranging from engineering to neuroscience~\cite{Pikovsky2003}. The paradigmatic Kuramoto model~\cite{Kuramoto1984} explains how an ensemble of phase oscillators can exhibit collective synchronization induced by identical all-to-all coupling, in spite of differences in their natural frequencies. Aside from realisations in biological systems~\cite{Jalife1987,Osaka2017}, synchronization of coupled oscillators finds technological application, e.g., in high-power laser diode arrays having high efficiency and low divergence~\cite{Jiang1993,Ruiz-Oliveras2009,Apollonov1998,Apollonov2014}.

Kuramoto~\cite{Kuramoto2002} discovered that the same type of non-linear interaction can lead to emergent phenomena for phase oscillators upon relaxing from a global identical coupling to a non-local coupling topology. These arrangements can be used to implement finite state machines~\cite{Orosz2009}, for example. However, they are also known to fail to synchronize completely, but rather to support coexistence of coherence and incoherence---later dubbed \emph{chimera states}~\cite{Abrams2004}---under specific conditions that are still being investigated. While the Kuramoto model is known to be analytically reducible with the Watanabe--Strogatz ansatz~\cite{Watanabe1994}, further results on generalized Kuramoto models with finite degrees of freedom or in the continuum limit gave insight into the relevant order parameters and their evolution in time~\cite{Pikovsky2008,Ott2008,Ott2009,Abrams2008}. Additional analyses were conducted showing that chimera states are robust to heterogeneities in natural frequencies~\cite{Montbrio2004,Laing20091,Laing20092}, coupling topologies~\cite{Shanahan2010,Kundu2018}, and noise~\cite{Laing2012}. Theoretical work~\cite{Panaggio2015,Omelchenko2018} also motivated successful experimental observations of such states in arrays of coupled chemical oscillators~\cite{Tinsley2012}, spatial light modulators~\cite{Hagerstrom2012}, and metronomes on swings~\cite{Martens2013}.

The study of the dynamics of micro-mechanical systems has undergone tremendous growth in recent years under the guise of optomechanics~\cite{Aspelmeyer2013}. The prototypical optomechanical system consists of a single mode of the electromagnetic radiation field, e.g., within a high-finesse optical cavity~\cite{Kippenberg2005}, interacting with the motion of a harmonic oscillator by means of the radiation pressure force~\cite{Marquardt2006}. The moving element variously takes the form of one of the end mirrors of a cavity~\cite{Groeblacher2009}, a semi-transparent membrane inside a cavity~\cite{Thompson2008}, one plate of a capacitor~\cite{Teufel2011}, a micro- or nano-particle~\cite{Millen2014,Delic2019} in a cavity, or the cavity itself in the case of micro-toroids supporting whispering gallery modes of the radiation field~\cite{Kippenberg2005}. The optomechanical interaction has been used to cool the motion of the mechanical system down to its ground state~\cite{Chan2011,Teufel2011}, generate quantum entanglement between mechanical oscillators~\cite{Ockeloen2018,Riedinger2018}, and produce proof-of-concept isolators and directional amplifiers for microwave radiation~\cite{Bernier2017,Malz2018,Barzanjeh2017}.

Recent work has started exploring the many-body dynamics of systems of coupled optomechanical networks, including the possibility of obtaining stronger coupling at the single-photon level~\cite{Xuereb2012,Piergentili2018} and their use to study synchronization phenomena~\cite{Heinrich2011,Lipson2012,Holmes2012,Lauter2015,Lipson2015}. Such systems may also find technological use; synchronized optomechanical arrays, for example, could act as high-power and low-noise on-chip frequency sources~\cite{Lipson2015}.

\begin{figure}
\includegraphics[width=0.475\textwidth,angle=0]{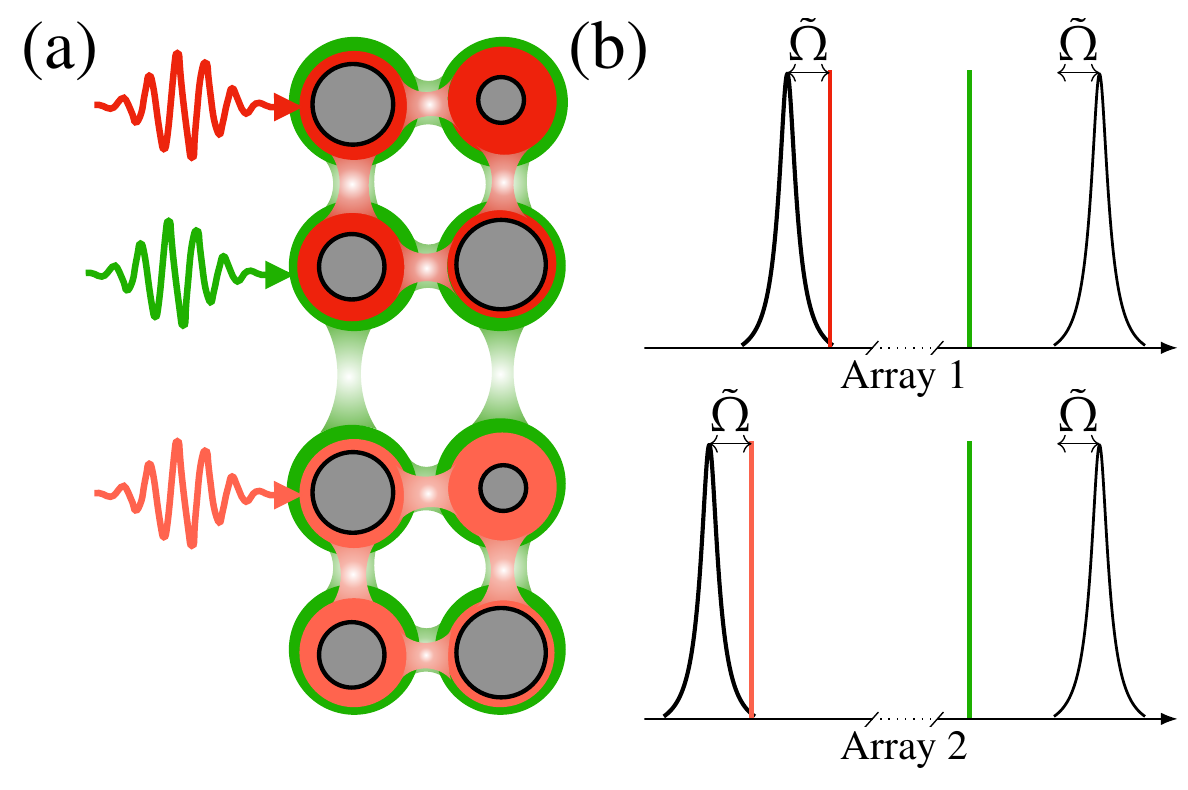}
\caption{A test-bed for investigating chimera-states: (a) Optomechanical micro-toroids are arranged in two identical arrays. Excitation of optical modes delocalised over each array causes self-sustained, synchronized mechanical oscillation at large optical powers. A third optical mode introduces spring-like coupling between both arrays allowing chimeras to emerge in the compound system. (b) Driving scheme: Each optical mode is driven coherently with a specific detuning. Further detail is given in the text.} 
\label{fig:Setup}
\end{figure}

The experimental observation of synchronization in optomechanical arrays~\cite{Lipson2015} raises the question of its robustness against disorder in the natural frequencies of the mechanical oscillators, as well as potential interactions of multiple arrays on one chip. To explore this question we employ a toy model of two identical arrays of optomechanical oscillators subject to global mechanical coupling within each, as well as with the other, array. Our theoretical analysis reveals how this disorder gives rise to chimera states close to the regions of parameter space where the arrays synchronize. Our work identifies limitations for the large scale integration of optomechanical arrays on a chip.

Following the introduction of our model, we describe the results of exhaustive numerical experiments that allow us to identify the region of parameter space where chimera states arise in our model. We then analyse the continuum limit to obtain analytical results, and conclude by discussing the implications of our results for applications of optomechanical arrays.

\emph{Model.}---We consider the collective dynamics of two identical optomechanical arrays, $\sigma=1,2$, each of which consists of $N$ mechanical modes coupled to one global laser-driven optical mode of amplitude $\alpha^{\sigma}$, which is described by the equations of motion~\cite{Heinrich2011,Lipson2012,Lipson2015}
\begin{subequations}
\label{eq:FullOMModel}
\begin{gather}
m\bigl(\ddot{x}_{i}^{\sigma}+\Gamma\dot{x}_{i}^{\sigma}+\Omega^{\sigma2}_{i}x_{i}^{\sigma}\bigr)=F_{i,\text{opt}}^{\sigma}(t)+\sum\limits_{\sigma'=1,2}F_{i,\text{m}}^{\sigma\sigma'},\ \text{and}\\
\dot{\alpha}^{\sigma}=\big[i\big(\Delta-\delta\omega_{\text{opt}}\big)-\tfrac{\kappa}{2}\big]\alpha^{\sigma}+\tfrac{\kappa}{2}\alpha_{\text{max}}.
\end{gather}
\end{subequations}
Here $x_j^{\sigma}$, $\Omega_i^\sigma$, $m$, and $\Gamma$ denote the displacement, natural frequency, effective mass, and damping rate of the mechanical modes; the last two are assumed to be identical to simplify the analysis but we have verified that our numerical results still hold for small variations of those. We assume that the arrays contain identical sets of oscillators, i.e., $\Omega_i^{1}=\Omega_i^{2}=:\Omega_i$ such that the two arrays are indistinguishable; chimera states are well-defined only in cases where the oscillator populations are identical. Each optical mode is characterised by its decay rate $\kappa$ and its detuning from the driving laser $\Delta=\omega_{\text{laser}}-\omega_{\text{opt}}$, which we assume are independent of $\sigma$. The optomechanical interaction shifts the resonance frequency by $\delta\omega_{\text{opt}}=-\sum Gx_j^{\sigma}$ as a result of the mechanical displacements, and imparts a force $F_{i,\text{opt}}^{\sigma}=\hbar G |\alpha^{\sigma}|^2$ on the mechanical modes. For blue detuning ($\Delta>0$) and large-enough optical power, there exists a Hopf bifurcation leading to synchronised self-oscillation~\cite{Marquardt2006,Heinrich2011,Lipson2015}.

The focus of this work is the analysis of additional mechanical coupling between the arrays and its effects on their synchronization. The mechanical coupling $F_{i,\text{m}}^{\sigma\sigma'}=\sum_j k_{ij}^{\sigma\sigma'}(x_j^{\sigma'}-x_i^{\sigma})$ is assumed to be global: $k_{ij}^{\sigma\sigma'}=(1-\delta_{ij}\delta_{\sigma\sigma'})\mu/N$, i.e., every pair of oscillators is coupled with strength $\mu/N$. Such global spring-like coupling was shown theoretically \cite{Heinrich2011} and realized experimentally for two oscillators \cite{Lipson2012} with an optomechanical coupling driven with a red detuned laser ($\Delta < 0$). \Fref{Setup} illustrates a schematic illustration of a realization of this model, i.e., a system of consisting of two arrays of micro-toroidal optomechanical systems that allows for optical modes delocalized \cite{Lipson2015} over either one array or both arrays. Driving each array with an optical mode delocalized over it with blue detuning allows the excitation of the mechanical oscillators and self-sustained synchronized oscillation at large powers. Driving another optical mode delocalized over both arrays with red detuning introduces global spring-like coupling adjustable via the input power and the detuning.

\begin{figure}[t]
\includegraphics[width=0.45\textwidth]{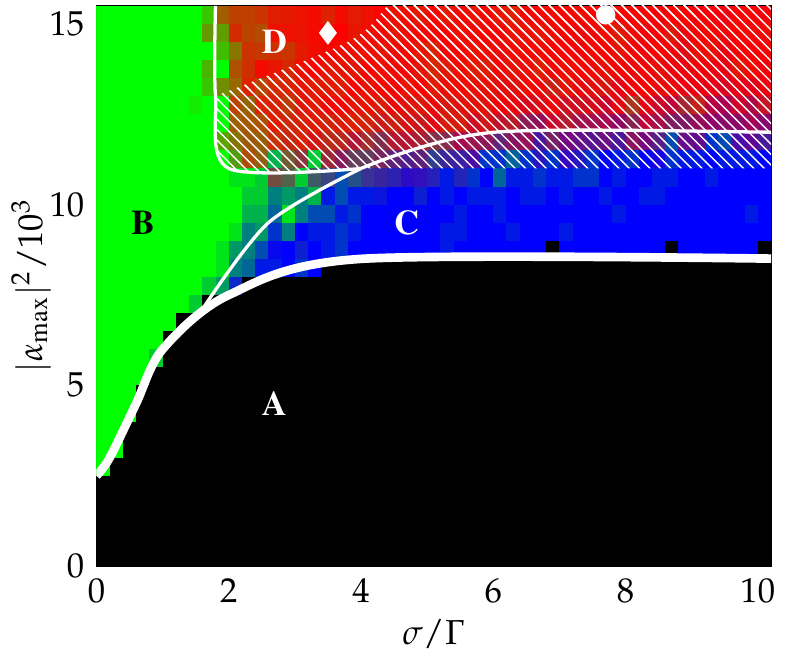}
\caption{(Color online) Schematic two dimensional parameter-space diagram of the synchronization behavior, plotted as a function of the standard deviation of the natural frequencies, $\sigma$ (horizontal axis), and the maximal number of photons $|\alpha_{\text{max}}|^2$ (vertical). Upon increasing the optical input power $P_{\alpha}\propto|\alpha_{\text{max}}|^2$ for distinct behaviors are observed depending on $\sigma$, characterizing the disorder: No self-sustained oscillation (Region A, black); self-sustained, synchronized oscillation at one frequency (Region B, green); self-sustained, unsychronised oscillation (Region C, blue); self-sustained, synchronized oscillation always attained at multiple frequencies (Region D, red). This plot can be seen as a projection of a three-dimensional parameter space, with the third axis corresponding to the mechanical coupling strength. In the hatched region, additional global mechanical coupling of two arrays leads to chimera states.}
\label{fig:Phasespace}
\end{figure}

The behavior of this system depends sensitively on the magnitude of the disorder in the frequencies of the mechanical elements, and on the input optical power. An overview of this behavior is depicted in \fref{Phasespace}, which is the main numerical result of this work. For weak disorder on the scale of a mechanical linewidth, we find that below a threshold input power $P_{\alpha} \propto|\alpha_{\text{max}}|^2$ there is no self-sustained oscillation (Region A). Above this threshold, we find synchronized oscillation of all mechanical oscillators for arbitrarily small additional mechanical spring-like interaction (Region B). The relative phase between the two arrays in the absence of the spring-like coupling is arbitrary and depends on the initial condition. In accordance with the analytical insight, one finds phase synchronization of the two arrays upon increasing the mechanical interaction.

\begin{figure}[t]
\begin{center}
\includegraphics[width=0.495\textwidth,angle=0]{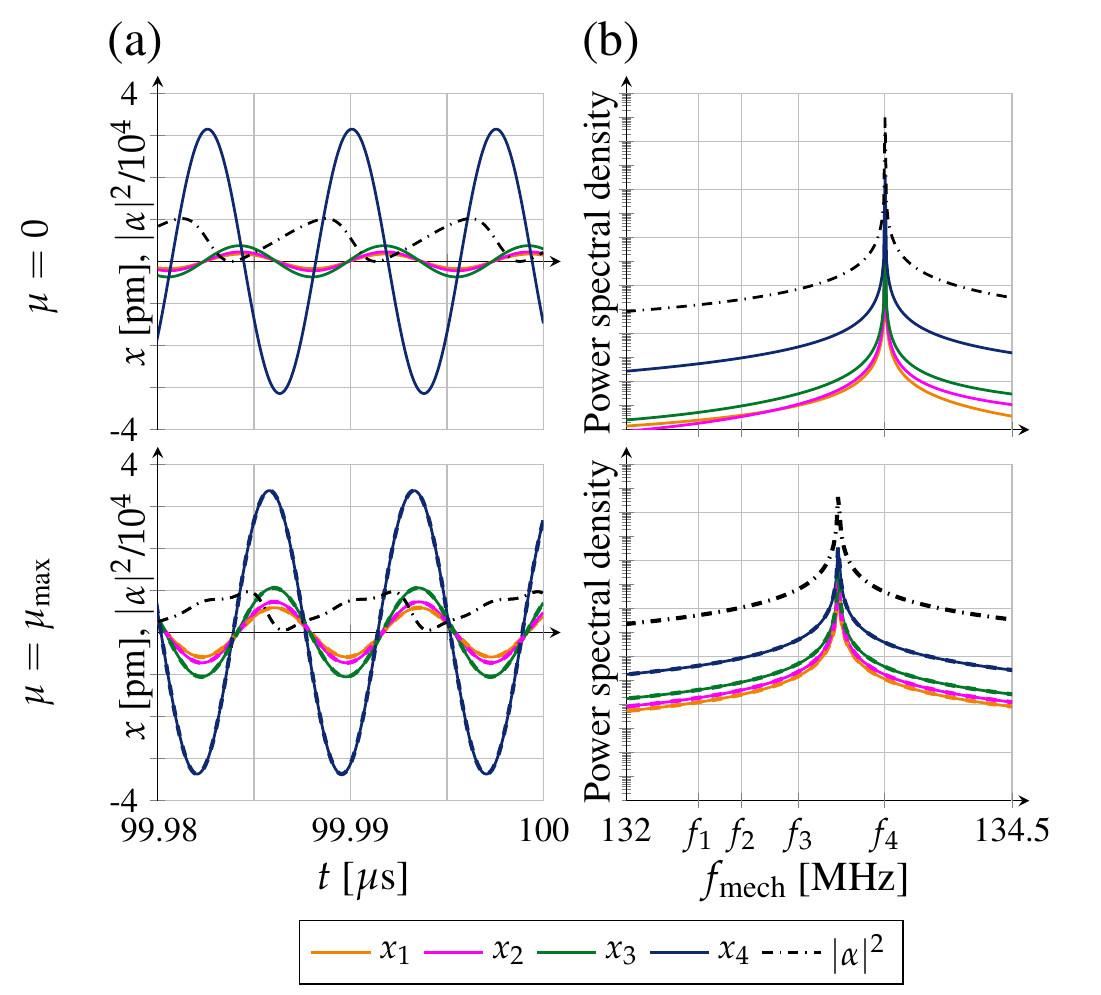}
\end{center}
\caption{(Color online) Numerical analysis of the behavior of an optomechanical array consisting of four oscillators and light field each in the absence of (upper row) and two with large mechanical coupling (lower row). The chosen parameters correspond to the white diamond in \fref{Phasespace}. (a)~Time evolution of the mechanical oscillators (solid orange, magenta, green, blue) and light field (dashed black) (b)~Mechanical power spectral densities. In both cases we find synchronization to one frequency; in the case of no mechanical coupling it is one of the natural frequencies of the oscillators (indicated by the vertical lines). In the large coupling case both arrays are synchronized in frequency and phase as can be seen by the evolution of the second array (thick dashed) plotted on top.}
\label{fig:Numerics1}
\end{figure}

For large-enough disorder, increasing the optical input power above the oscillation threshold leads to unsynchronized self-sustained oscillation of the arrays (Region C). Increasing the input power even further (Region D), one finds that both arrays always synchronize seperately to one of the natural frequencies $\Omega_i$ in absence of interaction of the two arrays. Introducing the mechanical interaction between the two arrays can drive one of the arrays out of the synchronized state while the other one is not affected. Since this coexistence of synchronization in one array and lack of sychronization in the other depends crucially on the mechanical coupling between the two arrays, and since the two arrays are identical, we can label these chimera states. Increasing the mechanical interaction between the arrays even further will eventually lead to in-phase or anti-phase synchronization of the two arrays.

\emph{Numerical results with small arrays.}---Under realistic circumstances, state-of-the-art optomechanical arrays consist of at most a few separate oscillators whose natural frequencies are spread beyond a linewidth (cf.\ Fig. S-5 in Ref.~\cite[SI]{Lipson2015}). To address the effect of additional mechanical coupling, we conducted numerical investigations of \erefs{FullOMModel} using parameters from Ref.~\cite{Lipson2015}. We consider two identical arrays, each consisting of four oscillators whose natural frequencies are centered around $\bar{\Omega}/2\pi= 133$\,MHz, and which have mechanical quality factor $\bar{\Omega}/\Gamma=1000$ and effective mass $m_{\text{eff}}=70$\,pg. The optical modes interact with each array with a coupling strength $G/2\pi=49$\,MHz/nm and have a decay rate $\kappa=\bar{\Omega}$. They are driven by a blue-detuned laser with $\Delta=\bar{\Omega}$. The mechanical coupling between the arrays is set to be global; we explore coupling strengths up to $|\mu_{\text{max}}|/{m\bar{\Omega}^2}=4.1 \times 10^{-3}$.

\begin{figure}
\begin{center}
\includegraphics[width=0.485\textwidth]{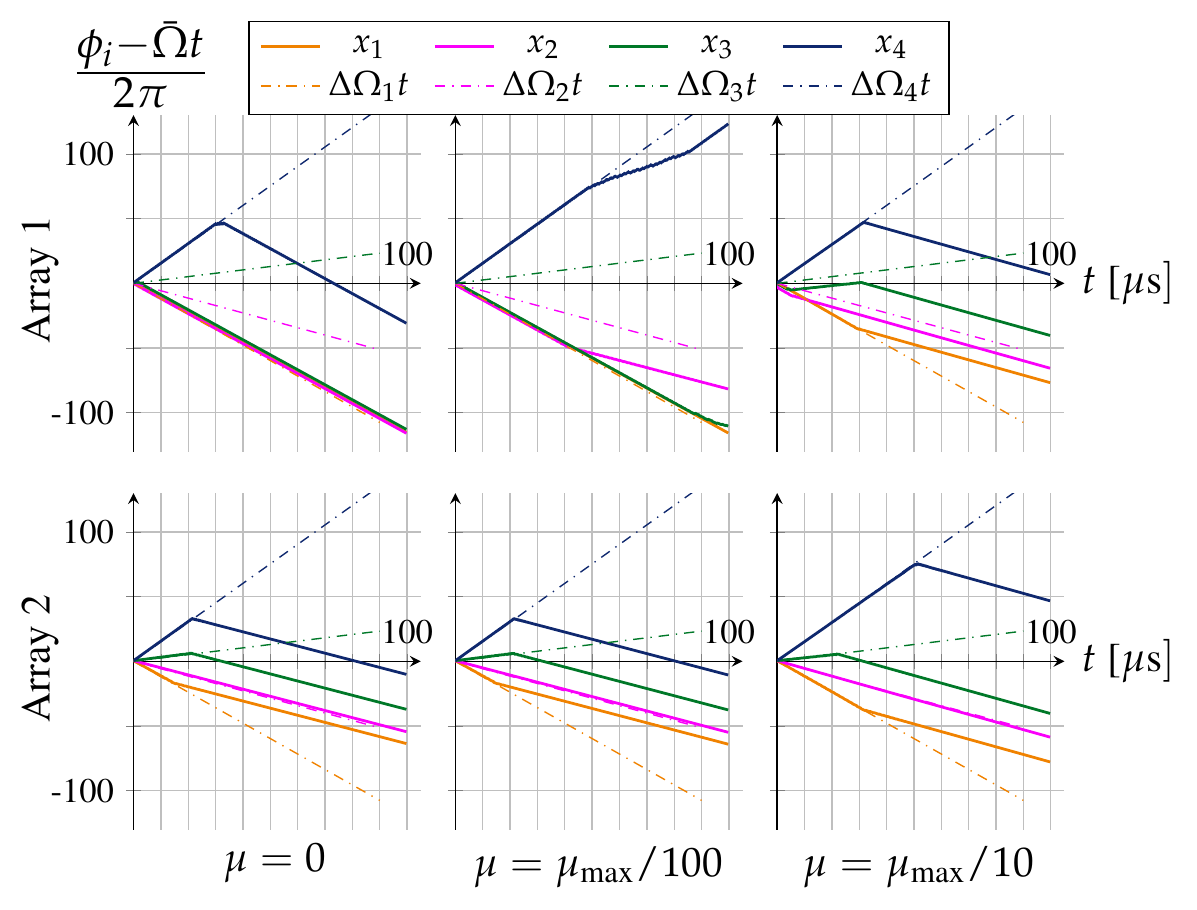}%
\end{center}
\caption{(Color online) Phase reconstruction of the mechanical motion (orange, magenta, green, blue) for both arrays (white circle in \fref{Phasespace}). The top (bottom) row refers to the first (second) array. In the absence of mechanical coupling (left), the two arrays synchronize, indicated by the same slope of all phases, seperately to different natural frequencies of the constituents. At intermediate mechanical coupling strengths (middle) the oscillators in the second array are synchronized to one frequency while in the first array the phases diverge. This is a chimera state, where the first array oscillates incoherently while the second array is synchronized. When mechanical coupling dominates (right), the two arrays synchronize to one another.}
\label{fig:Numerics2}
\end{figure}

We next turn to examples that illustrate the preceding discussion. \Fref{Numerics1} shows the behavior of the coupled arrays for zero (top) and dominant (bottom) mechanical coupling for parameters at the white diamond in \fref{Phasespace}. Without mechanical coupling, collective oscillation of each array takes place at one of the natural frequencies, showing the existence of coherence. Due to the large disorder, the phase difference between pairs of mechanical oscillators will be non-zero. When the mechanical coupling dominates, collective oscillation of the arrays take place at an arbitrary frequency with the two arrays oscillating in phase, in agreement with the analytical results presented below. 

\Fref{Numerics2} shows the phase reconstruction of the two arrays for different mechanical coupling strengths for parameters at the white circle in \fref{Phasespace}. The top (bottom) row represents the first (second) array, and the columns represent zero (left), intermediate (middle), and dominant (right) mechanical coupling. For zero mechanical coupling, the two arrays synchronize independently of one another to one of the natural frequencies of their oscillators. For large mechanical coupling, we again find that both arrays synchronize; this time, however, they both synchronize to the same frequency. However, at intermediate coupling strengths, we find that the phases of the first array diverge while the second array remains synchronized. This indicates that there is a stable coexistence of synchronization in one array and incoherent oscillation in the other, mediated by the mechanical coupling of the arrays.

\emph{Continuum limit.}---In the regime of self-sustained oscillations the light field can be adiabatically eliminated and the radiation pressure force $F_{\text{opt}}^{\sigma}(t)$ is then a periodic function with fundamental frequency $\bar{\Omega}$. Note that we can neglect the harmonics of the radiation pressure force and approximate $F_{\text{opt}}^{\sigma}(t)\approx F_0\sin(\bar{\Omega}t)$. By averaging over time and only considering the slow contributions one finds a generalised Kuramoto-like model for the phase evolution~\cite{Heinrich2011,Lauter2015}
\begin{align*}
\dot{\phi}_i^{\sigma}=&-\Omega_i+ K_i\sin(-\bar{\Omega}t+\tilde{\phi}_i^{\sigma}-\phi_i^{\sigma})\nonumber \\
&+\sum\limits_{\sigma'}\sum\limits_{j\in\sigma'}\frac{\xi^{\sigma\sigma'}_{ij}}{2}\cos(\phi_j^{\sigma'}-\phi_i^{\sigma})\nonumber\\
&+\sum\limits_{\sigma',\sigma''}\sum\limits_{\substack{k\in\sigma'' \\ j\in\sigma'}}\bigg\{\frac{\xi^{\sigma\sigma''}_{ik}\xi^{\sigma\sigma'}_{ij}}{4\Gamma}\sin(\phi_k^{\sigma''}+\phi_j^{\sigma'}-2\phi_i^{\sigma})+\nonumber \\
&\frac{\xi^{\sigma\sigma''}_{ik}\xi^{\sigma\sigma'}_{kj}}{4\Gamma}\bigl[\sin(2\phi_k^{\sigma''}-\phi_j^{\sigma'}-\phi_i^{\sigma})-\sin(\phi_j^{\sigma'}-\phi_i^{\sigma})\bigr]\bigg\},
\end{align*}
where we have defined $K_i=F_0/(2m\Omega_i\tilde{A}_i)$ and $\xi^{\sigma\sigma'}_{ij}=k^{\sigma\sigma'}_{ij}\tilde{A}_j/(m\Omega_i\tilde{A}_i)$. 

With the aim to obtain analytical insight using the Kuramoto-like model we follow the analysis of Ref.~\cite{Ott2008}. Thus, we require that the coupling constants $K_i$ can be considered the same and $\xi^{\sigma\sigma'}_{ij}$ to depend only on the array index $\sigma'$. Performing the continuum limit $N\rightarrow \infty$ requires the conservation of the number of oscillators for consistency. This results in continuity equations for the probability densities $f^{\sigma}(\Omega,\phi,t)=g(\Omega)\tilde{f}^{\sigma}(\phi,t)$ to find oscillators with natural frequency $\Omega$ to have phase $\phi$ at time $t$:
\begin{align}
\frac{\partial f^{\sigma}(\Omega,\phi,t)}{\partial t}+\frac{\partial}{\partial \phi}\bigl[f^{\sigma}(\Omega,\phi,t)v^{\sigma}(\Omega,\phi,t)\bigr]=0.
\label{eq:Continuity}
\end{align}
Following the method of Ref.~\cite{Ott2008}, we assume (i)~a Lorentzian natural frequency distribution $g(\Omega)=\{\pi[(\Omega-\bar{\Omega})^2+\varepsilon^2]\}^{-1}$, and (ii)~that the $\tilde{f}^{\sigma}(\theta,t)$ are periodic in $\theta$:
\begin{align}
\tilde{f}^{\sigma}(\theta,t)=\frac{1}{2\pi}\bigg\{1+\bigg[\sum\limits_{n=1}^{\infty}\tilde{f}^{\sigma}_n(t)\exp(in\theta)+c.c.\bigg]\bigg\},
\end{align}
with the Ott--Antonsen property $\tilde{f}^{\sigma}_n(t)=[a_{\sigma}(t)]^n$. This family of probability distributions contains the limiting cases of the uniform distribution for $\tilde{f}^{\sigma}_n(t)=0$, which signifies no knowledge about the phases, and $\delta(\theta-\Psi)$ for $\tilde{f}^{\sigma}_n(t)=e^{-in\Psi}$, signifying perfect synchronization of all phases to $\Psi$. Conveniently, $\tilde{f}^{\sigma}(\theta,t)$ converges for all $a_{\sigma}(t)=\rho_{\sigma}e^{-i\Psi_{\sigma}} \in \mathbb{C}$ with $\rho_{\sigma}\le1$ to
\begin{align}
\tilde{f}^{\sigma}(\theta,t)=\frac{1}{2\pi}\frac{(1-\rho_{\sigma})(1+\rho_{\sigma})}{(1-\rho_{\sigma})^2+4\rho_{\sigma}\sin^2\big[\frac{1}{2}(\theta-\Psi_\sigma)\big]}.
\end{align}
In absence of mechanical coupling, $\mu=0$, the phases $\Psi_{\sigma}$ decouple. The solution to the dynamics is $\Psi_{\sigma}=-\bar{\Omega} t+\bar{\phi}^{\sigma}$, and the corresponding stable fixed point is $\rho_{\sigma}=-\frac{2\varepsilon}{\Gamma}+\sqrt{1+(\frac{2\varepsilon}{\Gamma})^2}\xrightarrow{\varepsilon\ll\Gamma} 1$, i.e., perfect synchronization.

When the mechanical coupling $\mu$ is much larger than $\varepsilon$ and $\Gamma$, we can describe the dynamics of the system in terms of the phase difference $\Delta\Psi:=\Psi_1-\Psi_2$. We find the fixed points $(\rho_1,\rho_2,\Delta\Psi)=(1,1,n\pi)$ with $n\in\mathbb{Z}$, which describes two cases---either (i)~the two arrays synchronize perfectly, or (ii)~each array synchronizes seperately but in antiphase with the other array. This result generalizes the findings for two optomechanical oscillators in Ref.~\cite{Heinrich2011}, extending its applicability to two arrays of oscillators.

If all terms are relevant, analytical insight can be gained by assuming that array $1$, without loss of generality, is synchronized ($\rho_1=1$) and stays synchronized ($\dot{\rho}_1=0$). We obtain
\begin{align}
\rho_2=\sqrt{\frac{1-\varepsilon\Gamma(2m\bar{\Omega}/\mu)^2}{\cos(2\Delta\Psi)}}.
\label{eq:AnReq}
\end{align}
Since $\cos(2\Delta\Psi) \le 1$ for the relevant cases ($|\Delta\Psi|\ll\pi/4 \mod\pi$), we find that $\sqrt{1-\varepsilon\Gamma(2m\bar{\Omega}/\mu)^2}\le\rho_2\le1$. Importantly, this means that if there is no disorder ($\varepsilon=0$) there can be no chimera states ($\rho_2=1$). Chimera states \emph{require disorder} to exist in this system. Similarly, when the mechanical coupling dominates the disorder ($\mu^2\gg4\varepsilon\Gamma (m\bar{\Omega})^2$) the arrays must synchronize in phase or in antiphase.

\emph{Conclusions.}---Our investigation shows that highly complex nonlinear classical dynamics emerges in disorder strongly-driven optomechanical arrays. The fascinating pattern formation leading to the coexistence of coherence and incoherence in two interacting arrays is found to be enforced by the competition between two synchronization mechanisms. Since disorder in the natural frequencies of oscillators in realistic setups is of the order of a few linewidths, the physics we describe is of technological relevance. Our study further shows that this complex behavior is readily accessible to experiments.

In closing, we note that optics is not the only mechanism to couple the two arrays mechanically. Such interactions can also be introduced by strain coupling if the arrays are connected by a substrate, or sound waves if they are in close proximity to one another. Because of their generality, the effects we describe must be accounted for in systems with multiple non-linear oscillators in close proximity. If chimera states are to be avoided, our results imply a limit to the packing density of such arrays; if chimera states are to be sought, we have shown that a certain amount of disorder must be present.

\emph{Acknowledgments.}---This work is supported by the European Union's Horizon 2020 research and innovation programme under Grant Agreement No.\ 732894 (FET Proactive HOT).

\end{document}